\documentclass[a4paper,fleqn,usenatbib]{mnras}

\usepackage{newtxtext,newtxmath}
\usepackage{siunitx}
\usepackage{graphicx}
\usepackage{float}
\usepackage{subcaption}
\usepackage{calc}

\usepackage[T1]{fontenc}
\usepackage{ae,aecompl}

\usepackage[para,online,flushleft]{threeparttable}
\usepackage{tabularx}
\usepackage{lineno}
\usepackage{enumitem}
\usepackage{booktabs}

\usepackage{graphicx}	
\usepackage{amsmath}	

\title[Quiescent states of brightest blazars]{Multiwavelength Study of the Quiescent States of Six Brightest Flat Spectrum Radio Quasars detected by \textit{Fermi}-LAT}

\author[A. Roy et al.]{
Abhradeep Roy$^{1}$\thanks{E-mail: abhradeep.roy@tifr.res.in},
S. R. Patel$^{2}$,
A. Sarkar$^{1}$,
A. Chatterjee$^{1}$,
V. R. Chitnis$^{1}$
\\
$^{1}$Department of High Energy Physics, Tata Institute of Fundamental Research, Mumbai-400005, India\\
$^{2}$Deutsches Elektronen-Synchrotron, Platanenallee 6, D-15738 Zeuthen, Germany \\
}

\pubyear{2021}

\begin{document}
\label{firstpage}
\pagerange{\pageref{firstpage}--\pageref{lastpage}}
\maketitle

\begin{abstract}
The regular monitoring of flat-spectrum radio quasars (FSRQs) in $\gamma$-rays by \textit{Fermi}-LAT since past 12 years indicated six sources who exhibited extreme $\gamma$-ray outbursts crossing daily flux of 10$^{-5}$ photons cm$^{-2}$ s$^{-1}$. We obtained nearly-simultaneous multi-wavelength data of these sources in radio to $\gamma$-ray waveband from OVRO, Steward Observatory, SMARTS, \textit{Swift}-UVOT, \textit{Swift}-XRT and \textit{Fermi}-LAT. The time-averaged broadband Spectral Energy Distributions (SEDs) of these sources in quiescent states were studied to get an idea about the underlying baseline radiation processes. We modeled the SEDs using one-zone leptonic synchrotron and inverse-Compton emission scenario from broken power-law electron energy distribution inside a spherical plasma blob, relativistically moving down a conical jet. The model takes into account inverse-Compton scattering of externally and locally originated seed photons in the jet. The big blue bumps visible in quiescent state SEDs helped to estimate the accretion disk luminosities and central black hole masses. We found a correlation between the magnetic field inside the emission region and the ratio of emission region distance to disk luminosity, which implies that the magnetic field decreases with an increase in emission region distance and decrease in disk luminosity, suggesting a disk-jet connection. The high-energy index of the electron distribution was also found to be correlated with observed $\gamma$-ray luminosity as $\gamma$-rays are produced by high energy particles. In most cases, kinetic power carried by electrons can account for jet radiation power as jets become radiatively inefficient during quiescent states.
\end{abstract}




\section{Introduction}
\label{sec:intro}

\begin{table*}
	\centering
	\caption{ Details of studied blazars and selected epochs for quiescent state study.}
	\label{tab:sourcename}
	\begin{threeparttable}
	\begin{tabular}{lccccccc}
		\hline
		 \textbf{Name} & \textbf{R.A.} & \textbf{dec.} & \textbf{z} & \textbf{T$_{start}$} & \textbf{T$_{stop}$} & $L_{disk}$ & $M_{BH}$ \\
				       &     (deg)		  & (deg)         &            & DD-MM-YYYY (MJD) & DD-MM-YYYY (MJD) & ($10^{46}$ erg/s) & ($10^8$ $M_{\sun}$)\\
		\hline
		 CTA 102		&  338.158	  & 11.728        &  1.037     & 04-09-2011 (55808) & 18-10-2011 (55852) & 4.5 \tnote{a} & 8.5 \tnote{b}\\
		3C 279		&  194.045      & -5.786        &  0.536     & 14-01-2010 (55210) & 28-06-2010 (55375) & 0.1 \tnote{c} & 7.9 \tnote{d}\\
		3C 273		&  187.266      & 2.051         &  0.158     & 22-06-2012 (56100) & 11-08-2012 (56150) & 4.8 \tnote{a} & 21-30 \tnote{e} \\
		3C 454.3	&  343.493      & 16.149		  & 0.859	   & 19-05-2011 (55700) & 30-09-2012 (56200) & 6.75 \tnote{f} & 40 \tnote{g} \\
		PKS 1510-089  &  228.210	  & -9.106        & 0.361	   & 08-02-2011 (55600) & 19-05-2011 (55700) & 0.5 \tnote{h} & 13 \tnote{g}\\
		PKS B1222+216 &  186.226	  & 21.382		  & 0.432	   & 28-04-2015 (57140) & 07-06-2015 (57180) & 3.5 \tnote{i} &  6 \tnote{i}\\
		\hline
	\end{tabular}
	\begin{tablenotes}
		\item[a] \citet{10.1111/j.1365-2966.2009.15898.x}; \item[b] \citet{2014Natur.510..126Z}; \item[c] \citet{2015ApJ...803...15P}; \item[d] \citet{2009AA505601N}; \item[e] \citet{2005AA435811P}; \item[f] \citet{2011MNRAS.410..368B}; \item[g] \citet{2001MNRAS.327.1111G}; \item[h] \citet{2012ApJ...760...69N}; \item[i] \citet{2012MNRAS.424..393F}
	\end{tablenotes}
\end{threeparttable}
\end{table*}

Blazars form a subclass of Active Galactic Nuclei (AGN) whose relativistic particle outflows (jets) are aligned close to our line of sight \citep{1995PASP..107..803U}. Blazars are characterized by their significantly variable emission at all wavelengths across the electromagnetic spectrum ranging from radio frequencies to very high $\gamma$-ray energies. Blazars can be divided into two subclasses- BL Lacertae like object (BL Lac) and Flat Spectrum Radio Quasar (FSRQ). BL Lacs have almost featureless optical-UV spectra. Sometimes very weak narrow emission lines are present. FSRQs show stronger broad emission lines in optical spectra. The \textit{Fermi} Large Area Telescope (LAT) has detected 3137 blazars above 4$\sigma$ significance in 50 MeV to 1 TeV energy range (4FGL catalogue), out of which 22 BL Lacs and 43 FSRQs were identified in other wavebands also \citep{2020ApJS..247...33A}.

Blazars display flux variability on timescales of the order minutes to years when observed at $\gamma$-ray energies by \textit{Fermi}-LAT and ground-based Cherenkov telescopes \citep{2011ApJ...730L...8A, 2018ApJ...854L..26S}. From light travel-time argument, short variability timescales imply radiation from compact emission regions. Blazars are known for exhibiting occasional outbursts or flares. But, they spend most of their lifetime in the quiescent state. Studying these states allow us to probe the most common emission processes at work, estimating baseline parameters which can be useful in studying high flux activities. The direct thermal emission from accretion disks of FSRQs in quiescent states are generally visible in optical-UV waveband, but in other wavebands the jet emission dominates. Thus the study of quiescent states may help us to understand the underlying disk-jet connection and jet-empowering mechanisms.

We selected the six FSRQs (CTA 102, 3C 273, 3C 279, 3C 454.3, PKS 1510-089 \& PKS B1222+216) from the \textit{Fermi}-LAT monitored source list, also studied in \citet{2019ApJ...877...39M}, which are known for exhibiting brightest $\gamma$-ray flares with average daily fluxes over $10^{-5}$ cm$^{-2}$s$^{-1}$ within 1$\sigma$ statistical uncertainties above 100 MeV \citep{2019ApJ...877...39M}. Detailed multiwavelength studies on brightest $\gamma$-ray flares of all these sources have been done by various authors (CTA 102, \citet{2018ApJ...863..114G}; 3C 279, \citet{2016ApJ...832...17B}, \citet{10.1093/mnras/staa082}, \citet{Pittori_2018}; 3C 273, \citet{2013A&A...557A..71R}, \citet{2015A&A...576A.122E}; 3C 454.3, \citet{2016ApJ...826...54D}; PKS 1510-089, \citet{2014A&A...569A..46A}; PKS B1222+216, \citet{2014ApJ...786..157A}, \citet{10.1093/mnras/staa2958}). Except 3C 273, for the other five sources comparative study on quiescent states with flares have been done \citep{2017ApJ...851...72Z, 2012ApJ...754..114H, 2019arXiv190312381B, Acciari:2018ptm}. In this work, we  carried out a detailed study on quiescent state spectral energy distributions of these six brightest blazars. 

Broadband emission from blazars produces a typical double-humped SED extended from radio to $\gamma$-ray with low energy peak at IR to X-ray and the high energy peak at the $\gamma$-ray band. The blazar SEDs are often explained using leptonic and hadronic models which predict that the synchrotron emission from electrons rotating in a magnetic field of the jet produces the low energy peak. It is generally assumed that radiation is emitted by accelerated charged particles in plasma blobs moving down the jet at relativistic speeds. In leptonic processes, the high energy peak is produced due to inverse-Compton (IC) scattering of less energetic seed photons by the relativistic electrons present in the jet. Seed photons for IC scattering in leptonic model can be the synchrotron photons themselves (\underline{S}ynchrotron \underline{S}elf-\underline{C}ompton) or photons entering the emission region from accretion disk, broad-line region (BLR) and dusty torus region (\underline{E}xternal \underline{C}ompton). The spectral shape of the $\gamma$-rays generated by IC scattering is simply related to the energy distribution of the scatterer particles if the Comtponization lies in the Thomson regime depending on the energy of target seed photons. For Comptonization in the Klein-Nishina regime, the relation between $\gamma$-ray spectral shape and energy distribution of scatterer becomes more complicated. So, $\gamma$-ray spectral shape can act as an indicator of underlying particle acceleration mechanisms in the jet \citep{2020Ap&SS.365...33S}. Earlier studies suggest that underlying particle acceleration processes in blazar jet can be explained using various mechanisms like Fermi first order and second-order processes, magnetic reconnection, acceleration at recollimation shocks etc \citep{2020ApJ...890...56A, 2012arXiv1206.6147J, 2016A&A...592A..22H}.

In past, various kinds of emission scenarios have been applied to model the observed SEDs of these six sources in different states. In many past works, the blazar SEDs have been described using a hadronic model including photo-hadronic interactions and pair production process \citep{2012IJMPS...8..293S, 2011A&A...530A...4A, 2013A&A...555A..70T}. In hadronic models, the high energy hump in SED is explained as synchrotron emission from massive hadrons present in the jet, photo-hadronic interactions, pion decay, $e^-/e^+$ pair production and cascade emission. These hadronic models require very high magnetic fields ($\sim$100 G) and much harder proton spectrum, which seems to be unnatural \citep{2016ApJ...832...17B}. Some studies invoked multiple emission regions in order to get more satisfactory fit \citep{2019arXiv190312381B, 2011arXiv1110.6371B, 2019ApJ...883..137P, 2012PhRvD..86h5036T}. Time-dependant leptonic and hadronic models were used by some authors to calculate the radiating particle distribution solving the Fokker-Planck equation considering various acceleration processes and these models are helpful to understand the evolution of blazar SED during significant flux variations \citep{2018ApJ...861...31A, 2013MNRAS.431.2356Z, 2020ApJS..248....8D, 2015ApJ...809..171S}.  

A generalised continuous jet model was applied by \citet{2012MNRAS.423..756P, 2013MNRAS.429.1189P, 2013MNRAS.431.1840P, 2013MNRAS.436..304P} to explain blazar jet emission in a quiescent state. They have assumed a conical jet structure with a magnetic field dominated parabolic base region where particles get accelerated as suggested from VLBI image of M 87 jet. This model considers emission from the entire jet and so it can reproduce the observed radio emission when applied on different blazars' data. High energy emission has been explained by inverse-Compton scattering of seed photons coming from cosmic microwave background and narrow-line region. 

We used a leptonic one-zone synchrotron and inverse-Compton emission model to explain observed SEDs during quiescent source states. We imposed a strict conical jet model to constrain the emission region distance from the central black hole and found that the low $\gamma$-ray flux in the quiescent state can be explained by radiation from softer distribution of high energy particles inside large emission regions placed at distances estimated from a conical jet model. In \autoref{sec:obs} of this paper details of the instruments used and corresponding data analysis procedures are briefly discussed. Details of the SED model used for explaining observed data and the results obtained by modelling are mentioned in \autoref{sec:res}. Our inferences on physical processes inside the jet and correlations between different model parameters and observables are discussed in \autoref{sec:dis}. We listed the final conclusions in \autoref{sec:con}.

\begin{figure*}
	\centering
	\includegraphics[trim=40 0 0 0, clip, width=19cm, height=11cm]{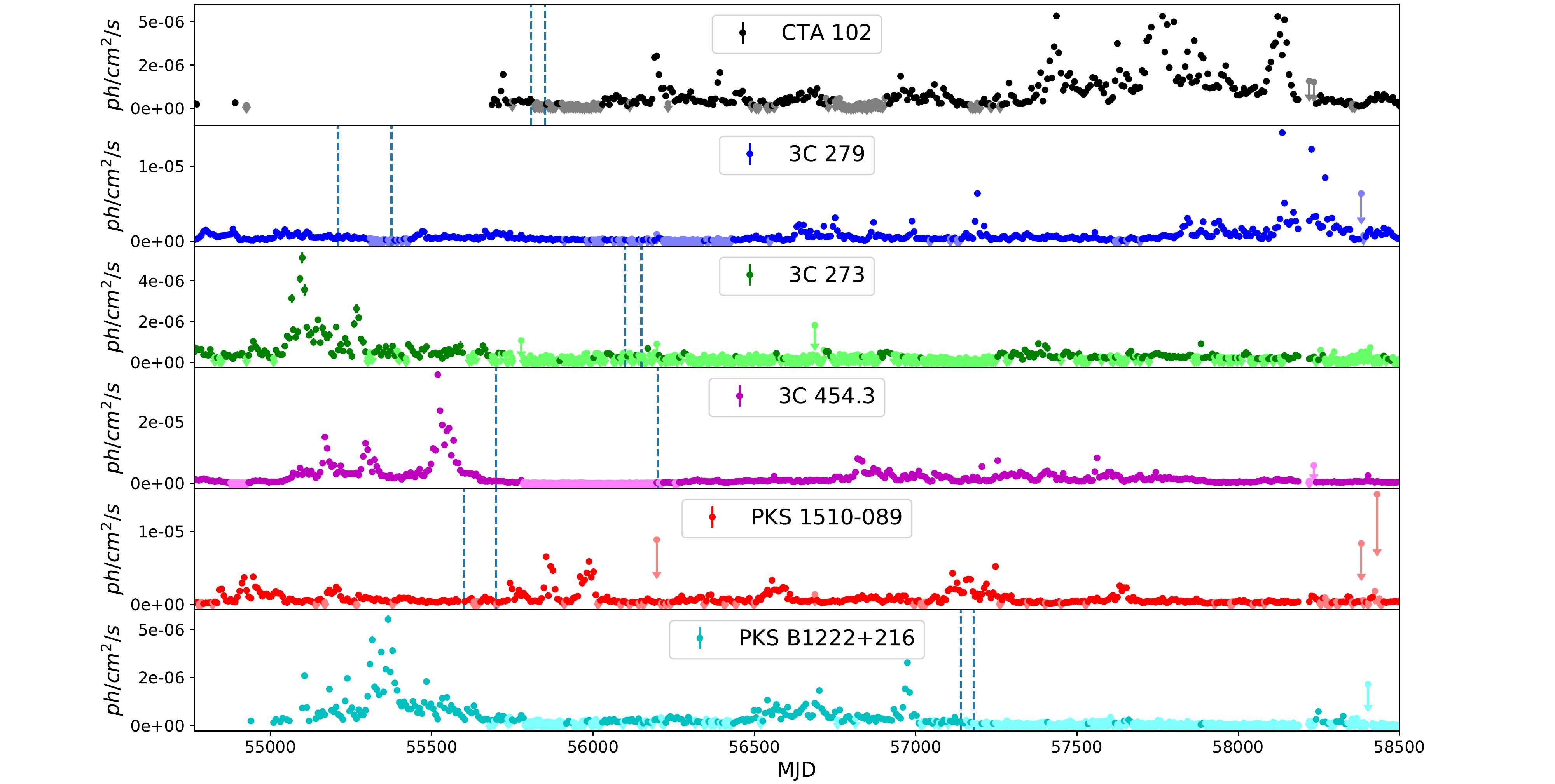}
	\caption{\textit{Fermi}-LAT weekly aperture photometry lightcurves of all sources. The selected quiescent state epochs are shown within vertical dashed blue lines.}
	\label{fig:alllc}
\end{figure*}

\section{Observations and Data Analysis} 
\label{sec:obs}

Some details of the selected sources and their quiescent epochs (T$_{start}$--T$_{stop}$) are listed in \autoref{tab:sourcename}. Publicly available radio lightcurves from Owens Valley Radio Observatory (OVRO), optical and infrared (IR) data from SPOL-CCD of Steward Observatory and Small and Moderate Aperture Research Telescope System (SMARTS) were used in this work. Optical-UV and X-ray data were obtained from UV/Optical Telescope (UVOT) and X-ray Telescope (XRT) onboard Neil Gehrels \textit{Swift} observatory respectively and analysed. High energy $\gamma$-ray data were taken from \textit{Fermi}-Large Area Telescope (LAT) and analysed. In this section, analysis procedures are discussed. Already published flux measurement data of all six sources were obtained from the archive of the Space Science Data Center (SSDC), a facility of the Italian Space Agency (ASI)\footnote{\url{https://www.ssdc.asi.it/}}.

\subsection{Radio Observations from OVRO}

Except PKSB 1222+216, all other sources are regularly monitored at 15 GHz by 40m telescope of OVRO, which is a part of \textit{Fermi} monitoring programme. Details of calibration, observations and data reduction technique of OVRO is given in \citet{2011ApJS..194...29R}. Each source is observed twice a week with a minimum flux density of about 4 mJy and typical 3\% uncertainties. Average flux density of each source within the epochs of our interest were computed in SED unit from OVRO lightcurves. Lightcurves of 3C 279, 3C 273 and 3C 454.3 are publicly available in the OVRO website\footnote{\url{https://www.astro.caltech.edu/ovroblazars/}}. CTA 102 and PKS 1510-089 lightcurves were kindly provided by OVRO collaboration. Observations of 3C 454.3 at 22 GHz and 37 GHz were obtained from 22 m radio telescope \citep{2000AstL...26..204N} of the Crimean Astrophysical Observatory (CrAO) and 14 m radio telescope \citep{1998A&AS..132..305T} of Aalto University Mets\"{a}hovi Radio Observatory respectively.

\subsection{Optical-NIR Obserations from SPOL-CCD of Steward Observatory and SMARTS}

SPOL-CCD imaging/Spectropolarimeter at Steward Observatory of the University of Arizona is also a part of Fermi multiwavelength support program and regularly monitors all the six blazars. Publicly available optical V and R-band photometric, polarimetric data and spectra were obtained from SPOL archive\footnote{\url{http://james.as.arizona.edu/~psmith/Fermi/}}. Details about these telescopes, instrumentation and data analysis procedures are discussed in \citet{2009arXiv0912.3621S}. Except CTA 102 and PKS B1222+216, observations of other sources  within selected epochs in optical (B, V, R) and near-infrared (J, K) wavebands were obtained from the SMARTS archive\footnote{\url{http://www.astro.yale.edu/smarts/glast/home.php}}. Details of SMARTS telescope design, detectors and data analysis procedures are given in \citet{2012ApJ...756...13B} and \citet{2012AJ....143..130B}. Observed optical-IR fluxes were corrected for Galactic reddening and extinction using an online tool\footnote{\url{https://irsa.ipac.caltech.edu/applications/DUST/}} contributed by IPAC at Caltech.

\subsection{Optical-UV Observations from \textit{Swift}-UVOT}

Optical-UV and X-ray data collected by space telescopes on-board Neil Gehrels \text{Swift}-mission were downloaded from NASA HEASARC archive\footnote{\url{https://heasarc.gsfc.nasa.gov/docs/cgro/db-perl/W3Browse/w3browse.pl}}. Optical-UV data were obtained from \textit{Swift}-UVOT \citep{2005SSRv..120...95R} which operates in imaging mode. It consists of three UV filters UM2, UW1, UW2 and three optical filters V, B, U. \textit{Swift}-UVOT and XRT data were analysed using tools bundled in HEASOFT (v 6.26) package. Data from all observations within our selected epochs were integrated using \textit{uvotimsum} tool. A circular source region of radius 10\arcsec was selected around the source position in the integrated images from each filter, and background was extracted from a circular region of radius 50\arcsec centred at the source-free region. Flux magnitudes were obtained using \textit{uvotsource} tool and then corrected for galactic extinction of E(B-V) according to \citet{1998ApJ...500..525S} for all filters using a web tool\footnote{\url{http://ned.ipac.caltech.edu/forms/calculator.html}}. Then the corrected observed magnitudes from all filters were converted into the fluxes in SED unit using zero-point magnitudes \citep{2008MNRAS.383..627P}. 

\subsection{X-Ray Observations from \textit{Swift}-XRT}

\textit{Swift}-XRT \citep{2005SSRv..120..165B} is a grazing incidence telescope which focusses soft X-rays in 0.3--8 keV energy range on CCD. XRT data were processed with XRTDAS (v3.5.0) software available within the HEASOFT package (v6.26). Data from all sources except PKS 1510-089 were recorded in photon counting mode. PKS 1510-089 data were recorded in windowed timing mode. Event files were cleaned and calibrated using \textit{xrtpipeline}(v0.13.5) and source spectra were obtained using \textit{xrtproducts} (v0.4.2). Source photons were extracted forming a circular region of 20 pixel-radius centred at RA and dec of the corresponding source. Background region was constructed as a circle having a radius of 40 pixels beside and completely detached from the source region. All XRT spectra within our time intervals of interest were combined using \textit{addspec} (v1.3.0) and then grouped to ensure a minimum of 20 counts in each energy bin using \textit{grppha} (v3.1.0). Standard auxiliary response files and response matrices were used during spectral fitting. Combined XRT spectra were fitted in \textit{xspec} (v12.0.1f) using power-law and log-parabola (whichever fits better) with the line of sight absorption in interstellar gas in terms of neutral hydrogen column density \citep{2005A&A...440..775K}.
The power-law model used is given as,  

\begin{equation}
\frac{dN}{dE} = k E^{-\Gamma}
\end{equation}
where $\Gamma$ is the spectral index. The log-parabola model is given as,

\begin{equation}
	 \frac{dN}{dE} = k \left( \frac{E}{E_b} \right) ^{-\alpha-\beta \log \left(E/E_b\right)} 	
\end{equation}
where $\alpha$ is the spectral index of at $E_b$ . During fitting $E_b$ was kept fixed at 1 keV. $\beta$ is curvature parameter.

\begin{table*}
	\centering
	\caption{15 GHz average radio fluxes calculated from OVRO lightcurves.}
	\label{tab:OVRO}
	\begin{tabular}{l|cccccc}
		\toprule
		Source & CTA 102 & 3C 279 & 3C 273 & 3C 454.3 & PKS 1510-089 & PKS B1222+216 \\
		\midrule
		flux (Jy) & 2.67$\pm$0.05 & 11.5$\pm$0.2 & 28.9$\pm$0.3 & 11.3$\pm$0.2 & 1.96$\pm$0.04 & -- \\
		\bottomrule
	\end{tabular}
\end{table*}

\begin{table*}
	\centering
	\caption{Reddening corrected average SPOL-CCD (R, V-filter) and SMARTS (K, J, R, V, B-filter) fluxes in SED unit.}
	\label{tab:SPOL}
	\begin{threeparttable}
	\begin{tabular}{lcccccccc}
		\toprule
		Source / Filters $\rightarrow$ & Unit & K & J & R & V & B & R  &  V  \\
		\hspace*{0.5cm} $\downarrow$   & (erg/cm$^2$/s) &   &   &   & & & (SPOL) &  (SPOL)  \\
		\midrule
		CTA 102       & $10^{-12}$ & -- & -- & -- & -- & -- & 3.33$\pm$0.04 & 3.52$\pm$0.02\\
		3C 279        & $10^{-12}$ & 7.345$\pm$0.005 & 3.647$\pm$0.003 & 2.286$\pm$0.002 & 1.983$\pm$0.002 & 1.601$\pm$0.003 & 1.924$\pm$0.002 & 1.853$\pm$0.002\\
		3C 273        & $10^{-10}$ & 1.255$\pm$0.001 & 0.974$\pm$0.001 & 1.276$\pm$0.001 & 1.766$\pm$0.001 & 1.987$\pm$0.001 & 1.33$\pm$0.02 & 1.61$\pm$0.03\\
		3C 454.3      & $10^{-12}$ & 3.5$\pm$0.4 & 5.3$\pm$0.2 & 5.9$\pm$0.2 & 7.3$\pm$0.1 & 6.7$\pm$0.1 & 6.049$\pm$0.007 & 6.78$\pm$0.01 \\
		PKS 1510-089  & $10^{-12}$ & 6.40$\pm$0.02 & 4.64$\pm$0.01 & 5.03$\pm$0.01 & 5.21$\pm$0.01 & 5.86$\pm$0.01 & 4.32$\pm$0.01 & 4.79$\pm$0.02\\
		PKS B1222+216 & $10^{-11}$ & -- & -- & -- & -- & -- & 1.25$\pm$0.06 & 1.5$\pm$0.1\\
	    \bottomrule
    \end{tabular}
	\begin{tablenotes}
		\textbf{Note:} All the SPOL-CCD and SMARTS observations within the selected epochs were averaged and corrected for reddening to obtain the fluxes mentioned above.
	\end{tablenotes}
	\end{threeparttable}
\end{table*}

\begin{table*}
	\centering
	\caption{Summary of the $\textit{Swift}$-UVOT data analysis.
		The fluxes of six UVOT filters are reported. Same observation IDs listed here were used for both UVOT and XRT analysis.
	}
	\label{tab:SwiftUVOT}
	\begin{threeparttable}
	\begin{tabular}{lccccccc}
		\toprule
		Source / Filters $\rightarrow$   & & V & B & U & W1 & M2 & W2 \\
		\hspace*{0.5cm} $\downarrow$ & Unit & & & & & & \\
		\midrule
		CTA 102 & $10^{-12}$ erg cm$^{-2}$ s$^{-1}$ & 4.1$\pm$0.1 & 4.03$\pm$0.09 & 5.09$\pm$ 0.08 & 5.8$\pm$0.1 & 6.6$\pm$0.1 & 5.10$\pm$0.09 \\
		3C 279 & $10^{-12}$ erg cm$^{-2}$ s$^{-1}$ & 2.1$\pm$0.1 & 1.86$\pm$0.07 & 1.38$\pm$0.04 & 1.17$\pm$0.03 & 1.21$\pm$0.04 & 1.11$\pm$0.03 \\
		3C 273 & $10^{-10}$ erg cm$^{-2}$ s$^{-1}$ & 1.71$\pm$0.04 & 1.98$\pm$0.06 & -- & 3.0$\pm$0.1 & 3.5$\pm$0.1 & 3.5$\pm$0.1 \\		
		3C 454.3 & $10^{-12}$ erg cm$^{-2}$ s$^{-1}$ & 9.2$\pm$0.5 & 8.4$\pm$0.3 & 9.3$\pm$0.3 & 8.8$\pm$0.3 & 9.9$\pm$0.4 & 7.7$\pm$0.2 \\
		PKS 1510-089 & $10^{-12}$ erg cm$^{-2}$ s$^{-1}$ & 6.2$\pm$0.5 & 7.0$\pm$0.3 & 7.5$\pm$0.3 & 5.8$\pm$0.3 & 7.1$\pm$0.3 & 6.7$\pm$0.2 \\
		PKS B1222+216 & $10^{-11}$ erg cm$^{-2}$ s$^{-1}$ & -- & -- & -- & -- & -- & 2.46$\pm$0.04 \\ 
		\bottomrule
	\end{tabular}
	\begin{tablenotes}
		\textbf{Note:} \textit{Swift} observation IDs used are listed below as range of starting observation ID to last observation ID within selected epochs. \\
		\textbf{CTA 102}: (00091094001--00091094018); \textbf{3C 279}: (00035019021--00035019037); \\ \textbf{3C 273}: (00035017114--00035017120); \textbf{3C 454.3}: (00035030204--00035030224); \\ \textbf{PKS 1510-089}: (00031173064--00031173071); \textbf{PKS B1222+216}: (00092193001--00092193007).
	\end{tablenotes}
	\end{threeparttable}
\end{table*}

\begin{table*}
	\centering
	\caption{Results from spectral fit of $\textit{Swift}$-XRT data.  
		For each of the flux states power law index ($\Gamma$), log parabola index ($\alpha$), curvature parameter ($\beta$), normalization factor of differential spectrum($k$), observed flux in 0.3 to 8.0 keV band (F$_{0.3-8 \ \text{keV}}$) and reduced chi square values ($\chi^2_r$) are reported.
	}
	\label{tab:SwiftXRT}
	\begin{tabular}{lcccccccc}
		\toprule
		Source  & n$_H$ & model &  $\Gamma$ & $\alpha$ & $\beta$  & $k$    & F$_{0.3-8 \ \text{keV}}$   & $\chi^2_r$      \\ 
		& (10$^{20}$ cm$^{-2}$)  &   &    & &       & (ph cm$^{-2}$ s$^{-1}$ keV$^{-1}$)      &  (erg cm$^{-2}$ s$^{-1}$) & \\
		\midrule
		CTA 102 & 4.81 & log parabola & -- & 1.03$\pm$0.05 & 0.28$\pm$0.08 & (4.05$\pm$0.09)$\times 10^{-4}$ & 3.81$\times 10^{-12}$ & 1.12\\
		3C 279 & 2.24 & log parabola & -- & 1.47$\pm$0.03 & 0.27$\pm$0.07 & (1.70$\pm$0.03)$\times 10^{-3}$ & 1.07$\times 10^{-11}$ & 1.05\\
		3C 273 & 1.69 & power law & 1.44$\pm$0.01 & -- & -- & (8.56$\pm$0.06)$\times 10^{-3}$ &  6.60$\times 10^{-11}$ & 1.24\\
		3C 454.3 & 6.78 & power law & 1.36$\pm$0.05 & -- & -- & (6.1$\pm$0.3)$\times 10^{-4}$ & 5.18$\times 10^{-12}$ & 0.98\\
		PKS 1510-089 & 7.13 & power law & 1.41$\pm$0.04 & -- & -- & (8.6$\pm$0.3)$\times 10^{-4}$ & 6.85$\times 10^{-12}$ & 1.07\\
		PKS B1222+216 & 1.72 & power law & 1.49$\pm$0.07 & -- & -- & (4.2$\pm$0.2)$\times 10^{-4}$ & 3.11$\times 10^{-12}$   & 1.52\\
		\bottomrule
	\end{tabular}
\end{table*}

\begin{table*}
	\centering
	\caption{Results from spectral fit of $\textit{Fermi}$-LAT data.
			For all sources the normalization parameter ($N_0$), spectral index ($\alpha$),
			curvature parameter of the spectrum ($\beta$), integrated flux in
			0.1--300 GeV (F$_{0.1-300 \ \text{GeV}}$) and test statistic value 
			of unbinned likelihood analysis (TS) are listed. 
	}
	\label{tab:Fermi}
	\begin{tabular}{lccccc}
		\toprule
		Source & $N_0$                                       & $\alpha $ & $\beta$ & F$_{0.1-300 \ \text{GeV}}$ & TS \\ 
		& (10$^{-10}$ ph cm$^{-2}$ s$^{-1}$ MeV$^{-1}$) &          &         & (10$^{-7}$ ph cm$^{2}$ s$^{-1}$) &    \\
		\midrule
		CTA 102 & 1.0$\pm$0.1 & 2.5$\pm$0.1 & 0.10$\pm$0.08 & 2.1$\pm$0.2 & 305.4 \\
		3C 279 & 0.84$\pm$0.05 & 2.46$\pm$0.05 & 0.11$\pm$0.04 & 2.0$\pm$0.1 & 1109.89 \\
		3C 273 & 2.0$\pm$0.3 & 2.9$\pm$0.2 & 0.2$\pm$0.2 & 1.9$\pm$0.2 & 158.73  \\
		3C 454.3 & 0.26$\pm$0.01  & 2.54$\pm$0.05 & 0.08$\pm$0.04 & 0.99$\pm$0.05 & 1091.34 \\
		PKS 1510-089 & 0.500$\pm$0.007 & 2.41$\pm$0.01 & 0.031$\pm$0.006 & 4.17$\pm$0.09 & 1909.65\\
		PKS B1222+216 & 0.59$\pm$0.07 & 2.6$\pm$0.1 & 0 & 1.3$\pm$0.2 & 143.16\\ 
		\bottomrule
	\end{tabular}
\end{table*}

\subsection{$\gamma$-Ray Observations from \textit{Fermi}-LAT}

\textit{Fermi} Large Area Telescope (LAT) is a pair production space telescope \citep{2009ApJ...697.1071A}. LAT has field of view $\sim$2.3 Sr and it covers 30 MeV to 1 TeV energy range. Data were analysed using standard software package \textit{fermitools}-v1.2.1 provided by \textit{Fermi}-LAT collaboration and user-contributed \textit{enrico} python script \citep{2013ICRC...33.2784S}. Data were collected in the energy range 0.1--300 GeV from \textit{Fermi}-LAT data archive\footnote{\url{https://fermi.gsfc.nasa.gov/ssc/data/access/}}. A circular region of interest (ROI) having a radius of 15$^{\circ}$ centred at the source was chosen for event reconstruction from the events belonging to SOURCE class. Events having zenith angle less than 95$^{\circ}$ were selected to get rid of $\gamma$-ray contribution from the Earth's albedo. Good time intervals were selected using a filter \lq\lq DATA\_QUAL$>$0"\&\& \lq\lq LAT\_CONFIG==1". The galactic diffused emission component gll\_iem\_v07.fits and an isotropic background emission model iso\_P8R3\_SOURCE\_V2\_v1.txt were used as background models. With an instrumental response function P8R3\_SOURCE\_V2, unbinned maximum likelihood analysis was carried out to obtain source spectrum. All the sources lying within ROI+10$^{\circ}$ radius around the source according to fourth \textit{Fermi}-LAT catalogue (4FGL) were included in the XML file. All parameters except the scaling factors were allowed to vary during the fitting process for sources within 5$^{\circ}$ from source position. The source-spectra were modelled using log-parabola as mentioned in 4FGL catalogue. The flux determination and spectral fitting were carried out by the likelihood analysis method using GTLIKE tool. Likelihood analysis was done iteratively by removing all sources having significances less than 1$\sigma$ after each fitting process. The entire energy range of each source was divided into few bins for obtaining flux points in SED unit. For flux points having test statistics less than 9 (i.e. $<$3$\sigma$ significance), flux upperlimits were estimated at 95\% confidence level using profile likelihood method. 
\textit{Fermi} unfiltered aperture photometry light-curves were obtained from their website\footnote{\url{https://fermi.gsfc.nasa.gov/ssc/data/access/lat/msl\_lc}}.

\section{Results}
\label{sec:res}

\textit{Fermi}-LAT weekly aperture photometry lightcurves were used to select quiescent states for all the sources (\autoref{fig:alllc}). The quiescent states were selected such that there are a bunch of \textit{Swift}-XRT observations present during these epochs so that we can make nearly simultaneous SEDs. It was also checked whether the sources were in quiescent states in radio, optical-UV, X-ray and $\gamma$-ray wavebands altogether. Average radio fluxes obtained from OVRO are listed in \autoref{tab:OVRO}. Average fluxes in optical-IR waveband obtained from Steward Observatory and SMARTS are shown in \autoref{tab:SPOL}.
Results of spectral analysis for combined UVOT and XRT observations from all sources within selected epochs are listed in \autoref{tab:SwiftUVOT} and \autoref{tab:SwiftXRT} respectively. Spectral fit results of \textit{Fermi}-LAT data are listed in \autoref{tab:Fermi}. 
We have made time-averaged broadband spectral energy distributions of all the six FSRQs during their quiescent states and modeled them with one-zone leptonic scenario using numerical code \lq Jetset' provided by Andrea Tramacere \citep{2006A&A...448..861M, 2009A&A...501..879T, 2011ApJ...739...66T, 2020ascl.soft09001T}. The model parameters for all sources were studied and compared in detail.

\subsection{The SED model}

SEDs of the six FSRQs were fitted with leptonic one-zone synchrotron and inverse Compton model. The inverse Compton process takes into account the respective contributions from synchrotron photons generated inside the jet (synchrotron self-Compton or SSC), external photon field coming directly from the accretion disk, and reprocessed disk photon field coming from BLR and dusty torus (external Compton or EC).\par
In this model, the source of broadband emission is assumed to be a spherical plasma blob of radius \lq $R$' located at a distance \lq $d$' from the central supermassive black hole of mass \lq $M_{BH}$'. The blob is moving relativistically down the jet with a bulk Lorentz factor \lq $\Gamma$'. Magnetic field (B) inside the blob is assumed to be same and isotropic everywhere.  \par
A non-thermal population of electrons having energy distribution of broken power-law shape was considered.

\begin{equation} \label{eq:3}
n\left(\gamma \right) = \left\{ \begin{array}{cc} 
k \gamma^{-p} & \hspace{5mm} \gamma \leq \gamma_{break} \\
k \gamma_{break}^{(p_1-p)} \gamma^{-p_1} & \hspace{5mm} \gamma > \gamma_{break} \\
\end{array} \right.
\end{equation}
This electron population gets cooled by synchrotron emission due to interaction with the magnetic field inside emission blob and generates the low energy hump in SED. The synchrotron photons get Compton upscattered due to collision with the synchrotron emitting relativistic electron population (SSC). Photons emitted by the accretion disk can directly enter the emission region or can get reprocessed from BLR and dusty torus and enter the emission blob and get Compton upscattered (EC). Thus, SSC and EC process produces high energy hump in the SED. As the emission blob moves down the jet, the radiation gets Doppler boosted in observer's frame by a factor $\delta$ along our line of sight given as,
$\delta = [\Gamma\left(1-\beta cos\theta\right)]^{-1}$,
where $\Gamma$ is the bulk Lorentz factor of the emission blob and $\theta$ is the angle between the jet axis and our line of sight. These values for all sources were taken from literature. The radius of the emission region was constrained using the light travel-time argument as,

\begin{equation}
R \leq \frac{c \delta t_{var}}{2\left(1+z\right)}
\end{equation}
where $t_{var}$ is the variability timescale.  
Distance of the emission region from the central engine, \lq $d$', was constrained using conical jet model so that the emission blob fills the entire jet cross-section.

Using accretion disk luminosity \lq $L_d$', BLR distance ($R_{BLR}$) and dusty torus distance ($R_{DT}$) from central engine were calculated by applying scaling relations derived from reverberation mapping technique \citep{10.1111/j.1365-2966.2009.15898.x}. 

\begin{equation}
R_{BLR} = 10^{17} \times (L_d/10^{45})^{1/2}
\end{equation}
\begin{equation}
R_{DT} = 2.5 \times 10^{18} \times (L_d/10^{45})^{1/2}
\end{equation}
Assuming that BLR is a thin spherical shell of ionized gases, the inner and outer radii of BLR were selected around $R_{BLR}$ such that, $R_{BLR}$ = ($R_{BLR}^{in}$+$R_{BLR}^{out}$)/2 and ($R_{BLR}^{out}$-$R_{BLR}^{in}$) = $2\times10^{16}$ cm. The temperature of the dusty torus was kept at typical 1000 K for all sources. Multi-temperature blackbody type accretion disk model was used where the temperature of a portion on the disk depends on its distance from the core as,

\begin{equation}
T^4\left(r \right) = \frac{3 R_S L_d}{16 \epsilon \pi \sigma_{SB} r^3}\left(1 - \sqrt{\frac{3 R_S}{r}}\right)
\end{equation}
where $\epsilon$ is the accretion efficiency which was set to 0.08 and $\sigma_{SB}$ is the Stefan-Boltzmann constant. The accretion disk was assumed to be extended from $3 R_S$ distance from central engine to a radius of $500 R_S$. $\tau_{BLR}$ and $\tau_{DT}$ represents the fraction of accretion disk emission intercepted and reprocessed by BLR and dusty torus respectively. Photon fields from BLR and dusty torus get Doppler boosted by a factor $\sim$$\Gamma^2$ in the blob-comoving frame till the blob is inside BLR and torus. This model first calculates the energy densities and luminosities in the blob comoving frame. Then, luminosities are converted into flux in observer frame by calculating luminosity distance ($D_L$) from the given redshift using cosmological model with $\Omega_\Lambda = 0.685$, $\Omega_M = 0.315$ and Hubble's constant H$_0$= 67.3 km s$^{-1}$ Mpc$^{-1}$ \citep{2014A&A...571A..16P}.

\subsection{SED modelling approach}

(i) Radio data do not constrain the emission model. Radio flux point did not get fitted by synchrotron radiation component because this model takes into account self-absorption of synchrotron emission at low frequencies below $\sim$$10^{12}$ Hz by synchrotron emitting electrons in the compact emission region (synchrotron self-absorption). This implies that other extended regions in the jet significantly contribute to radio emission where cross-section for synchrotron self-absorption process is less. \\
(ii) Optical-UV flux points form a small bump in the SED of most FSRQs. This bump is interpreted as direct thermal emission from the accretion disk. Varying disk luminosity ($L_d$) and $M_{BH}$ within 10\% around the values quoted in literature (see \autoref{tab:sourcename}) the height and peak position of the thermal radiation component can be fitted. Good UVOT observations are required for this process. \\
(iii) Available data do not give enough coverage to infer peak frequencies of synchrotron and inverse Compton radiation component. Due to the lack of simultaneous mm-IR observations, it is not possible to constrain synchrotron peak position.\\
(iv) The high energy slope of the electron energy distribution ($p_1$) was constrained using the slope of \textit{Fermi}-LAT flux points. Value of $p_1$ greater than 3 implies that the peak of the synchrotron and inverse Compton radiation is produced by the electrons at the break of their energy distribution (i.e. $\gamma_{peak} = \gamma_{break}$). Sometimes, for low $\gamma_{peak}$ values, the synchrotron peaks are produced below the self-absorption frequencies. In these cases, self-absorption frequencies become peak synchrotron frequencies.\\
(v) The jet viewing angle ($\theta$) and bulk Lorentz factor ($\Gamma$) were quoted from literature. \\
(vi) The magnetic field ($B$) and electron density ($N_e$) were constrained by fitting the synchrotron component to the IR observations from SMARTS. But in the absence of IR data in some cases, low energy end of the X-ray spectrum is fitted with significant contribution from SSC component (e.g. CTA 102, PKS 1510-089) which help to constrain magnetic field. For blazars having low disk luminosities (e.g. 3C 279), synchrotron emission dominates over thermal disk emission in the optical-UV region and thus $B$ and $N_e$ can be constrained. Sometimes even when a big blue bump is visible, the shape of the optical-UV spectrum ensures some contribution from synchrotron emission along with thermal disk emission (e.g. CTA 102, 3C 454.3, PKS B1222+216). \\
(vii) Other parameters like $p$, $\gamma_{min}$ and $\gamma_{max}$ were varied within feasible values to fit the SEDs by eye estimation due to sparse data sampling.

\begin{figure*}
	\centering
	\includegraphics[trim=120 0 60 0, clip, width=19cm, height=12cm]{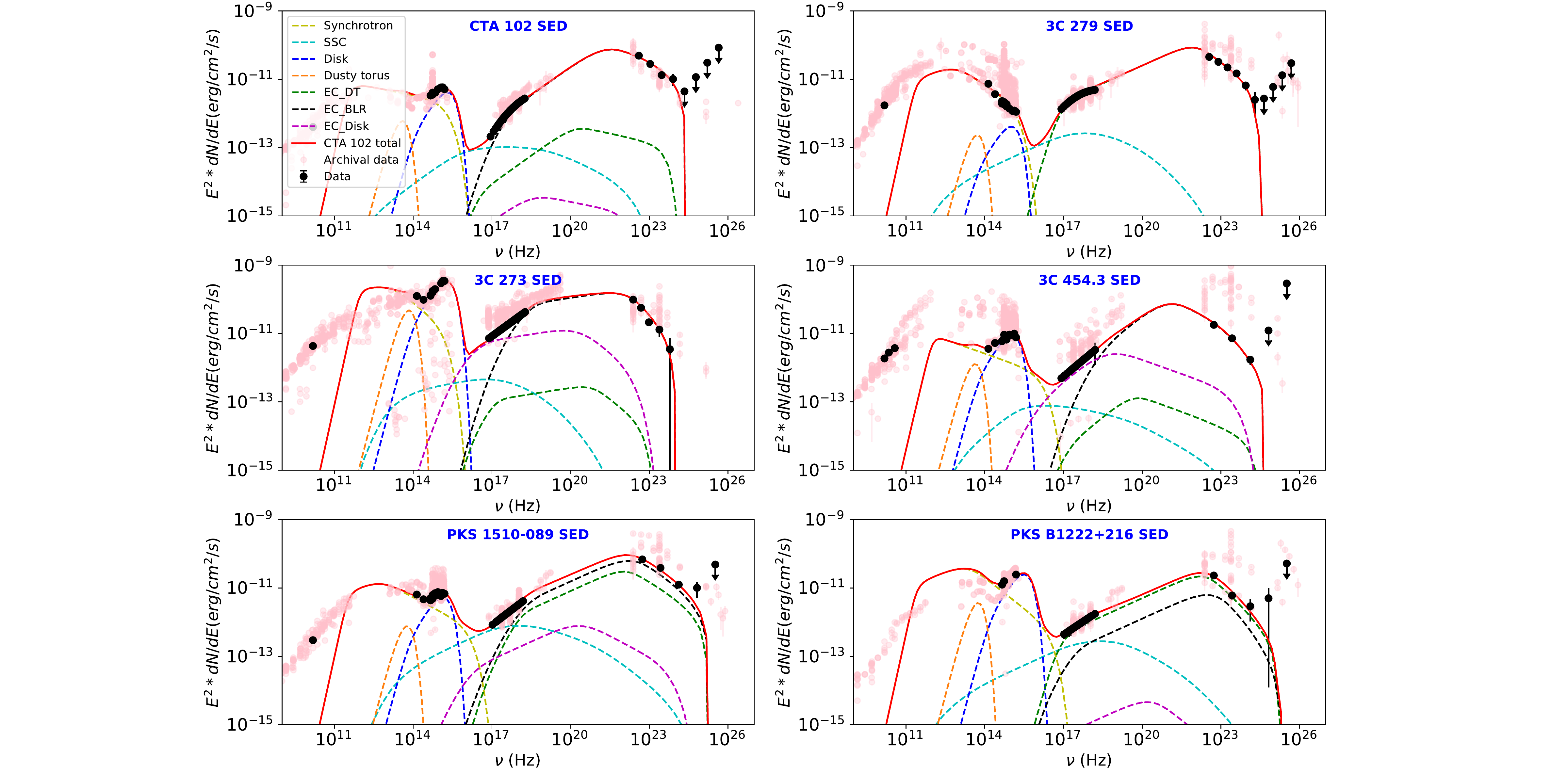}
	\caption{Fitted SEDs of the six brightest blazars in quiescent state. Black points represent analysed flux measurements within the selected epochs and pink points represent archived data from earlier studies. Yellow, blue, orange, cyan, black, green and magenta dashed lines indicate synchrotron, disk thermal, dusty-torus thermal, SSC, EC$_{BLR}$, EC$_{DT}$ and EC$_{Disk}$ emission component respectively. The solid red line represent total broadband emission.}
	\label{fig:allsed}
\end{figure*}

\begin{table*}
	\centering
	\caption{The input parameters of the model used to reproduce the 
		observed SEDs.}
	\label{tab:SEDParameter}
	\begin{threeparttable}
	\begin{tabular}{llcccccc}
		\toprule
		\textbf{Sr.} & \textbf{Parameters}   & \textbf{CTA 102}   & \textbf{3C 279} & \textbf{3C 273} & \textbf{3C 454.3}	& \textbf{PKS} & \textbf{PKS} \\
		\textbf{No.} &&&&&& \textbf{1510-089} & \textbf{1222+216} \\
		\midrule
		1. & $R$ (10$^{16}$ cm)      & 4.9             &  6.0     & 2.6	  & 1.3	     & 2.0	    & 6.0\\
		2. & $\theta$ (degree)        & 3.7 \tnote{a}      &  2.4 \tnote{a}    & 3.0 \tnote{b}    & 1.3	\tnote{a}     & 2.2	\tnote{*}    & 2.5 \tnote{d}\\   
		3. & $\Gamma$                 & 15.5 \tnote{a}    &  20.9 \tnote{a}  & 9.0 \tnote{b}    & 19.9 \tnote{a}	 & 20.0 \tnote{c}	    & 23.0 \tnote{*}\\  
		4. & $\delta$                 & 15.5 \tnote{*}           &  23.67 \tnote{*}   & 14.7 \tnote{*}   & 33.0 \tnote{*}	 & 25.0	\tnote{c}    & 23.0 \tnote{d} \\  
		5. & $d$ (10$^{17}$ cm)       & 7.5            &  15.0    & 2.0	  & 2.5	     & 4.0 	    & 14.0\\
		6. & B (G)                    & 2.3            &  0.9     & 12.0 	  & 8.0	     & 0.6	    & 1.35\\	
		7. & $\gamma_{min}$           & 1              &  1.1       & 1.7     &	1.4      & 3	    & 1\\
		8. & $\gamma_{max}$           & 4000           &  4500    & 1200    & 5000	 & 15000	& 15000\\
		9. & $\gamma_{break}$         & 72             &  340     & 140     & 25     & 350	    & 480\\
		10.& $N_e$ (cm$^{-3}$)        & 4200           &  3000    & 5000    & 3000    & 10000	    & 1550\\ 
		11.& $p$			          & 1.9            &  2.3     & 2.75    & 2.0	     & 2.3	    & 2.3\\
		12.& $p_1$                     & 3.3            &  4.0     & 4.2     & 3.54     & 3.8	    & 4.2\\
		\hline
		& & \multicolumn{4}{c}{Parameters external to the jet} & & \\
		\hline
		13.& $L_{d}$             (10$^{46}$ erg s$^{-1}$) & 5.0      &  0.1      & 4.8    & 6.75    & 0.5	& 3.6\\ 
		14.& M$_{BH}$  (M$_{\sun}$)                       & 8.5E+8   &  7.9E+8   & 3.0E+9 & 4.0E+9	& 1.3E+9	& 6E+8\\ 
		15.& $R_{BLR}^{in}$ (10$^{17}$ cm)                & 7.0      &  0.9      & 6.9    & 8.0	    & 2.1	& 5.9\\
		16.& $R_{BLR}^{out}$ (10$^{17}$ cm)               & 7.2      &  1.1      & 7.1    &	8.2     & 2.3	& 6.1\\
		17.& $\tau_{BLR}$                                 & 0.15     &  0.1      & 0.1    & 0.1	    & 0.1	& 0.1\\
		18.& T$_{DT}$ (K)                                 & 1000     &  1000     & 1000   & 1000	& 1000	& 1000\\
		29.&$R_{DT}$(10$^{18}$ cm)                        & 16.7     &  2.5      & 17.3   &	20.5    & 5.6   & 15\\
		20.&$\tau_{DT}$					                  & 0.1		 &  0.39	 & 0.1    & 0.1	    & 0.1	& 0.1\\	
		\bottomrule  
	\end{tabular}
	\begin{tablenotes}
		\item[a] \citet{2009AA498723H}; \item[b] \citet{Paliya_2017}; \item[c] \citet{Acciari:2018ptm}; \item[d] \citet{2014ApJ79661K}; \item[*] Computed using, $\delta = [\Gamma\left(1-\beta cos\theta\right)]^{-1}$
	\end{tablenotes}
\end{threeparttable}
\end{table*}
\begin{table*}
	\centering
	\caption{Energy densities, powers and mass accretion rates calculated from the parameters of SED modelling.}
	\label{tab:EnergyPower}
	\begin{tabular}{llcccccc}
		\toprule
		\textbf{Sr.} & \textbf{Parameters}   & \textbf{CTA 102}   & \textbf{3C 279} & \textbf{3C 273} & \textbf{3C 454.3}	& \textbf{PKS} & \textbf{PKS} \\
		\textbf{No.} &&&&&& \textbf{1510-089} & \textbf{1222+216} \\
		\midrule
		1. &  $U_e$	    (erg/cc)  & 1.46e-02   &  9.92e-03   &  1.58e-02  & 1.26e-02    & 8.52e-02	& 4.75e-03\\
		2. &  $U_B$     (erg/cc)  & 4.07e-01   &  3.22e-02   &  5.73      & 2.55	    & 1.43e-02  & 7.25e-02\\
		3. & $\eta$ ($U_e$/$U_B$) &  0.036     &  0.308      &  0.003     & 0.005	    & 5.958     & 0.065   \\
		4. & $P_{rad}$  (erg/s)   & 3.32e+45   &  2.70e+44   &  7.63e+43  & 1.42e+44	& 8.65e+43	& 1.28e+44\\
		5. & $P_{kin}$  (erg/s)   & 4.74e+46   &  7.30e+46   &  3.35e+46  & 1.61e+46	& 2.41e+46	& 5.56e+46\\
		6. & $P_{e}$    (erg/s)   & 7.92e+44   &  1.47e+45   &  8.17e+43  & 7.92e+43	& 1.28e+45	& 8.53e+44\\
		7. & $P_{p}$    (erg/s)   & 2.45e+46   &  6.68e+46   &  3.88e+45  & 2.84e+45 	& 2.26e+46	& 4.18e+46\\
		8. & $P_B$      (erg/s)   & 2.21e+46   &  1.30e+46   &  2.95e+46  & 1.61e+46	& 2.15e+44	& 8.74e+44\\
		9. & $P_{jet}$  (erg/s)   & 4.74e+46   &  5.63e+46   &  3.35e+46  & 1.90e+46    & 2.41e+46  & 5.03e+46\\
		10. & $L_{Edd}$ (erg/s)	  &	1.07e+47   &  9.95e+46	 &	3.78e+47  & 5.04e+47    & 1.64e+47	& 7.56e+46\\
		11. & $L_d/L_{Edd}$       & 0.47       &  0.01       & 0.13       & 0.13        & 0.03      & 0.48\\
		12. & $\dot{M}_{accr}$ ($M_{\sun}/yr$) & 10.95       & 0.22       & 10.51       & 14.78     & 1.10   & 7.88 \\
		13. & $\dot{M}_{out}$ ($M_{\sun}/yr$)  & 0.028       & 0.056      & 0.007       & 0.003     & 0.020  & 0.032 \\
		14. & $\dot{M}_{out}/\dot{M}_{accr}$   & 0.003       & 0.256      & 0.001       & 0.0002     & 0.018  & 0.004 \\
		15. & $\dot{M}_{accr}/\dot{M}_{Edd}$   & 5.84        & 0.13       & 1.59        & 1.67      & 0.38   & 5.95 \\
		\bottomrule        
	\end{tabular}
\end{table*}

\subsection{SED fit results}

SEDs of six sources fitted with a one-zone leptonic model are shown in \autoref{fig:allsed} and the sets of fit parameters are listed in \autoref{tab:SEDParameter}. 
The trends shown by the archived data helped to model the SEDs. When all UVOT filters were not available, the underlying trend of archived data helped us to decide whether optical-UV were forming a part of the disk thermal emission bump. These trends also helped in modelling the height and shape of the synchrotron peak. In some earlier studies of FSRQs, the X-ray spectra were fitted by SSC component and $\gamma$-ray spectra were fitted with EC component \citep{2015MNRAS.454.1310Y, 2020A&A...635A..25S, 2014ApJ...790...45P, 2012ApJ...754..114H, 2015ApJ...803...15P}. Though hard X-ray data were not available in our case, the underlying trends in archived data implied that soft X-ray and $\gamma$-ray spectra form the rising and decaying part of a single broad EC component. For this purpose, the $\gamma_{min}$ values were kept low (within 1--10). Except for 3C 279 and PKS B1222+216, the single broad EC components could not describe the low energy ends of the soft X-ray spectra. The contributions from EC emissions of Doppler de-boosted disk radiation ($EC_{disk}$) were required in 3C 273 and 3C 454.3 SED, and SSC contribution was required in CTA 102 and PKS 1510-089 SED. Synchrotron emission flux from 3C 273 is the highest and that of 3C 454.3 is the lowest. PKS B1222+216 has the lowest EC flux and 3C 273 has the highest.

Derived black hole masses ($M_{BH}$) were found to be in the range $(7-40) \times 10^8 M_{\sun}$ and Disk luminosities were found to be in the range $(1-68)\times10^{45}$ erg s$^{-1}$ for our selected sources. All these values are consistant with values mentioned in literature. In the case of 3C 279, the optical-UV data did not indicate a dominance of disk thermal emission over synchrotron radiation. So, in this case, $L_d$ and $M_{BH}$ values were quoted from literature. Derived disk luminosities are in between 1 and 48 per cent of the Eddington luminosity ($L_{Edd}$).

For four out of six sources, inverse Compton emission was found to be dominated by $EC_{BLR}$ component, though the emission region is inside BLR in only two sources (3C 273 and 3C 454.3). $\tau_{BLR}$ values were found to be 10\% (except 15\% for CTA 102) and $\tau_{DT}$ values are 10\% (except 38\% for 3C 279).

Magnetic fields were found in range ($0.6-12$) G. Electron densities of ($10^3-10^4$) cm$^{-3}$ in the emission region indicate the presence of very powerful jets in FSRQs. Low energy slopes of the particle distribution ($p$) were found to be greater than 2 for four sources except CTA 102 and 3C 454.3. Values of $p_1$ were found to be greater than 3 and break energies were found in the range ($30-500$).

\subsection{Power estimation}

Relativistic jet contains power in form of radiation ($P_{rad}$), electrons ($P_e$), protons ($P_p$) and magnetic field ($P_B$). Powers are estimated mainly using,

\begin{equation}
P_i \simeq \pi R^2 \Gamma^2 \beta c U_i
\end{equation}
where $U_i$ is the energy density of the $i^{th}$ power carrying component in the comoving frame. Thus, power carried by the relativistic electron population is given as,

\begin{equation}
P_e \simeq \pi R^2 \Gamma^2 \beta \langle \gamma \rangle N_e m_e c^3
\end{equation}
and power carried by \lq cold' protons (as they are massive they do not get accelerated enough to move relativistically) are estimated as,

\begin{equation}
P_p \simeq \pi R^2 \Gamma^2 \beta  N_p m_p c^3
\end{equation}
where $\langle \gamma \rangle$ is the average Lorentz factor of relativistic electrons, $m_e$ and $m_p$ are rest masses of electron and proton respectively, $N_e$ and $N_p$ are electron and proton density respectively inside the emission blob. $N_e$ goes as input parameter in the model. We assumed presence of one proton per ten electrons inside blob \citep{2014Natur.515..376G} to calculate $P_p$. The power carried by magnetic field as Poynting flux is given as,

\begin{equation}
P_B \simeq \frac{1}{8}R^2 \Gamma^2 \beta c B^2
\end{equation}

\begin{figure}
	\centering
	\includegraphics[trim=150 0 50 0, width=8.7cm, height=5cm]{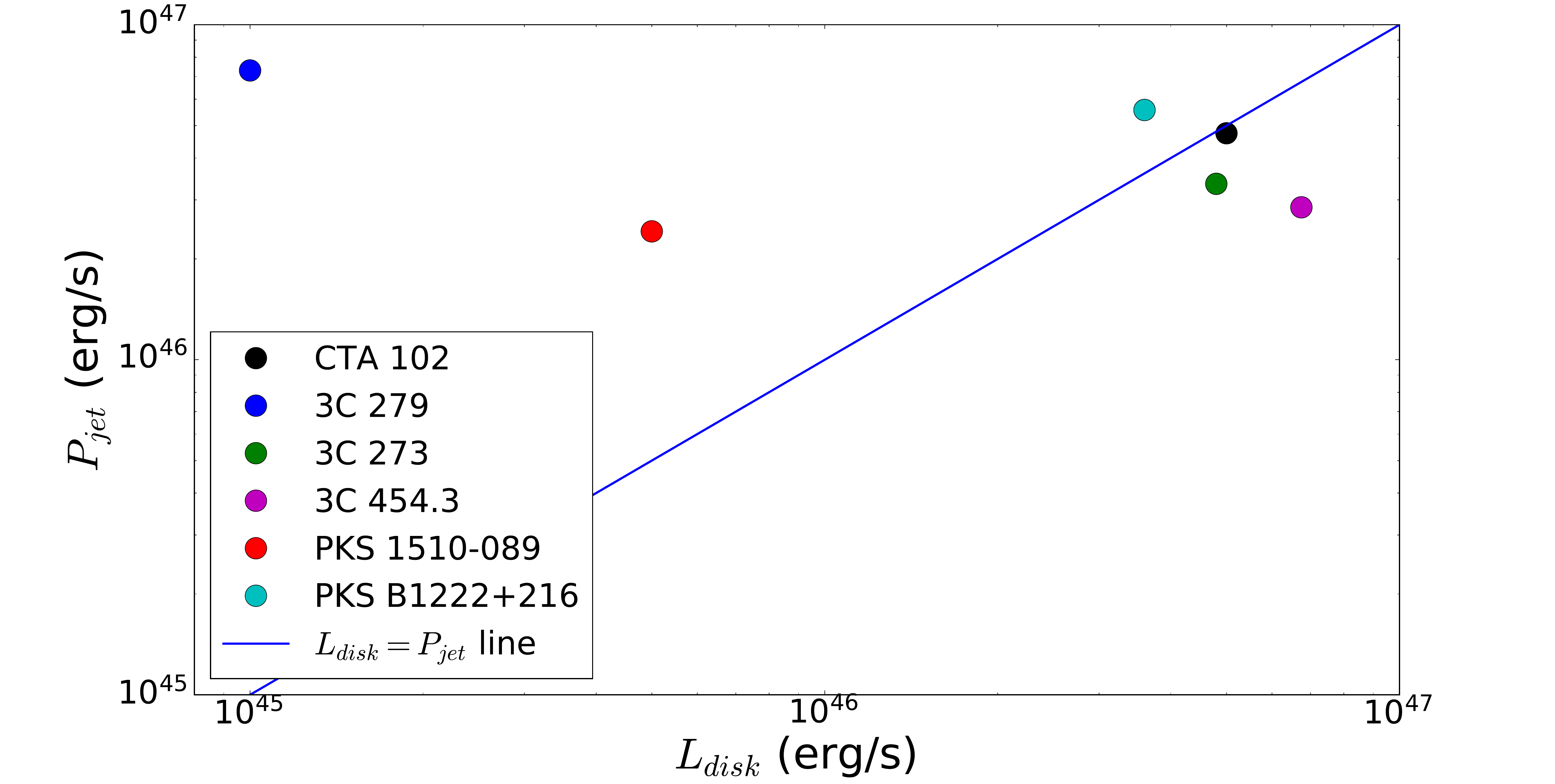}
	\caption{Disk luminosity vs jet power plot.}
	\label{fig:ldvslj}
\end{figure}

Total jet power ($P_{jet}$) is the sum of powers carried by electrons, protons and magnetic field. Power radiated by the jet ($P_{rad}$) can be restated using radiation energy density in comoving frame ($U_{r} = L'/(4\pi R^2 c)$) as,

\begin{equation}
P_{rad} = L'\frac{\Gamma^2}{4} = L\frac{\Gamma^2}{4\delta^4}
\end{equation}
where $L'$ is the total non-thermal radiation luminosity in comoving frame and $L$ is the total luminosity in observer's frame.
Four out of six FSRQs have jet powers in quiescent state greater than or almost equal to disk luminosities, which indicates that FSRQs have very powerful jets (\autoref{fig:ldvslj}). Calculated powers are listed in \autoref{tab:EnergyPower}. Total jet powers of all sources were found to be less than the Eddington luminosities.

\subsection{Estimation of mass outflow rate}

Mass accretion rate can be estimated from disk luminosity as,

\begin{equation}
\dot{M}_{accr} = \frac{L_d}{\epsilon c^2}	
\end{equation}
Mass outflow rate can be estimated as,

\begin{equation}
\dot{M}_{out} = \frac{P_{jet}}{\Gamma c^2}
\end{equation}	
Except 3C 279, all other sources require more than 1 $M_{\sun}$ yr$^{-1}$ mass accretion rate. But the mass outflow rate is $\sim 0.003-0.3$ times of accretion rate. This can be understood as $\epsilon/\Gamma << 1$ and $P_{jet}$ and $L_d$ have same orders of magnitude.

\begin{figure}
	\centering
	\includegraphics[trim=140 0 60 0, width=8.7cm, height=5cm]{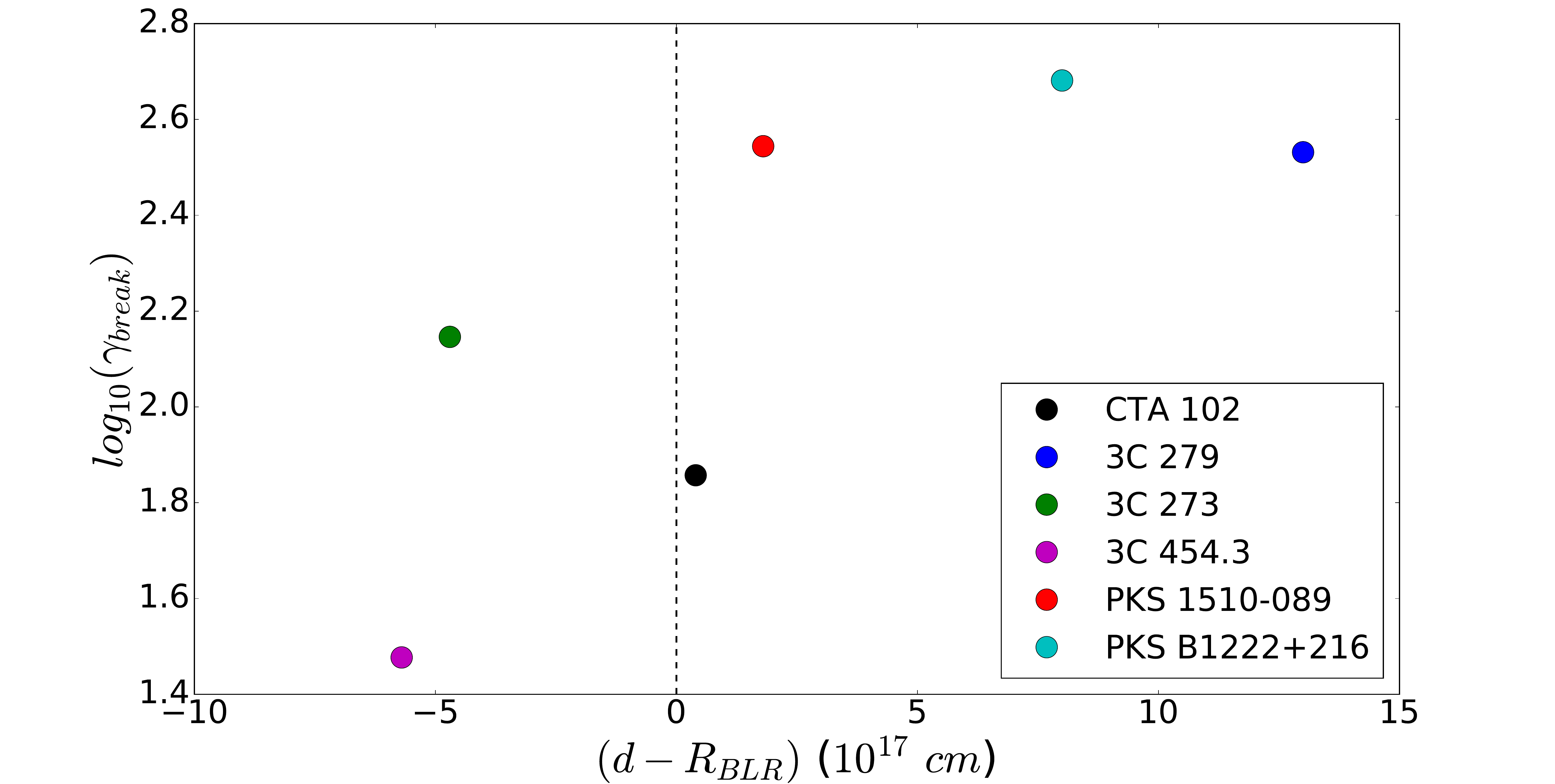}
	\caption{Dependence of break energy on difference of emission region distance and BLR distance.}
	\label{fig:gvsdrblr}
\end{figure}
\begin{figure}
	\centering
	\includegraphics[trim=140 0 50 0, width=9cm, height=5cm]{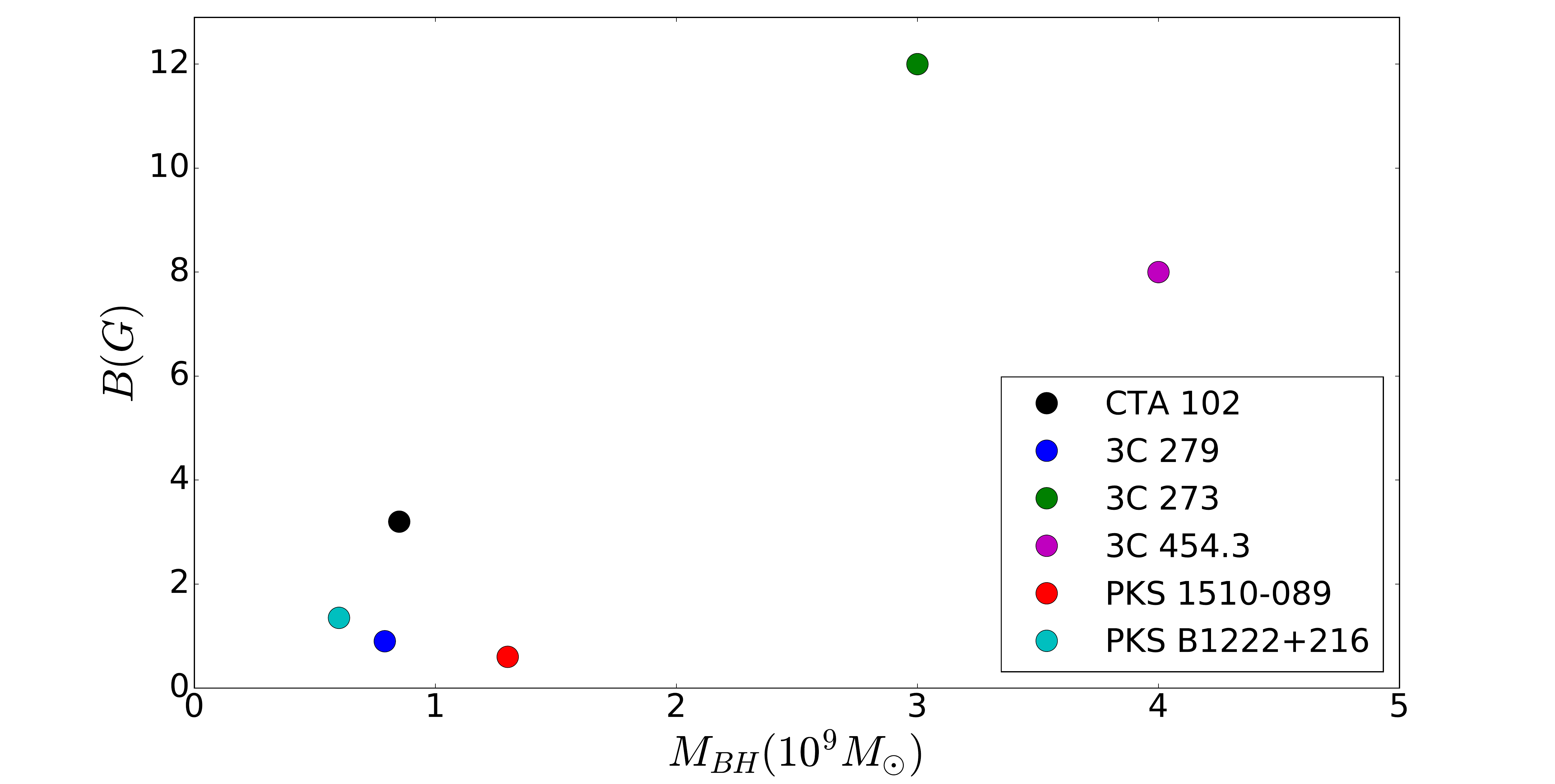}
	\caption{Dependence of magnetic field (B) in jet on central black hole mass ($M_{BH}$).}
	\label{fig:mbhvsb}
\end{figure}
\begin{figure}
	\centering
	\includegraphics[trim=40 0 10 0, width=8cm, height=5cm]{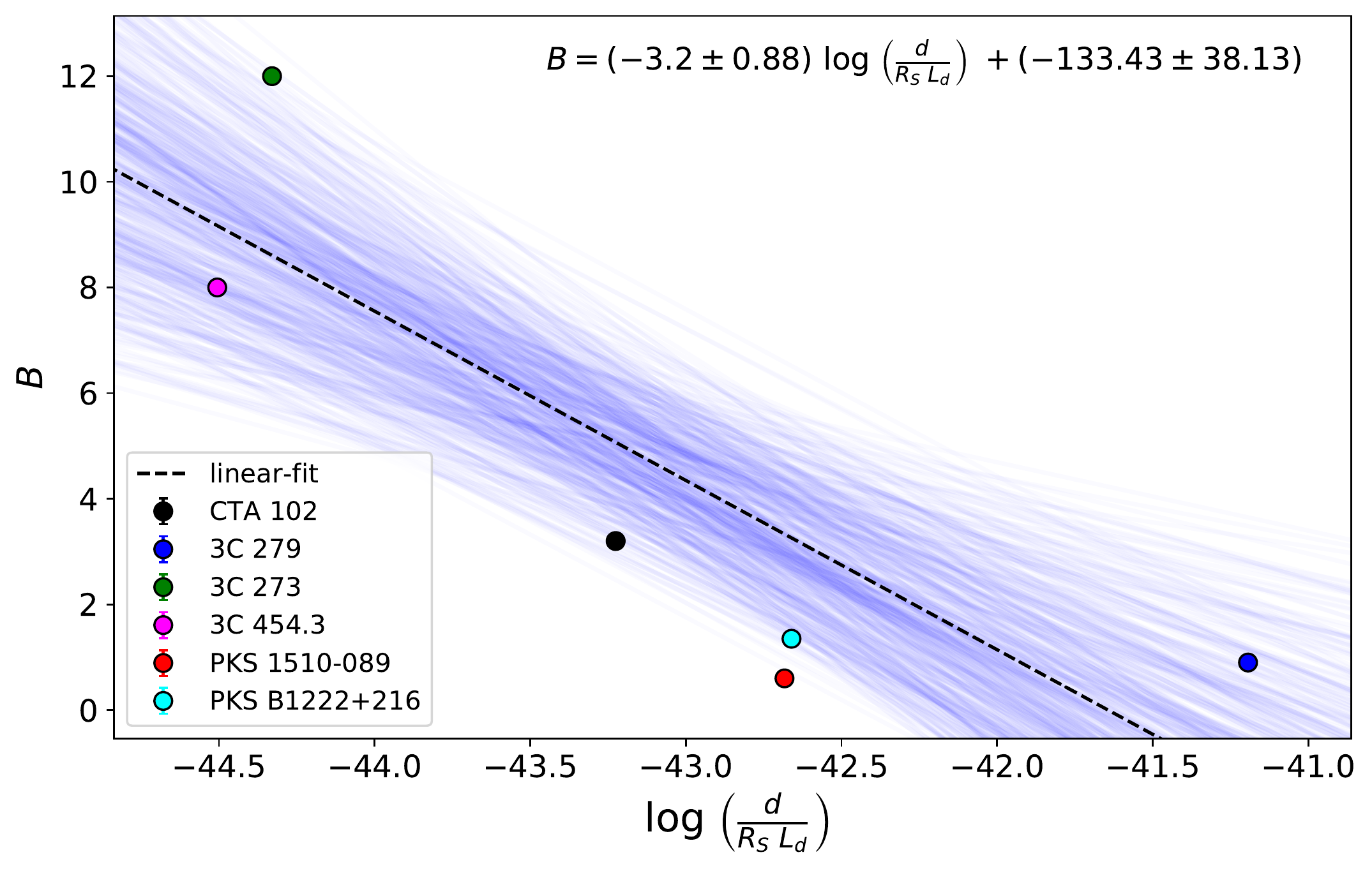}
	\caption{Dependence of magnetic field on distance of emission region ($d$), black hole mass ($M_{BH}$) and disk luminosity ($L_d$) fitted with bootstrapped linear regression. Blue lines are bootstrapped regression lines and the black dotted line is the average of them.}
	\label{fig:bvslogdld}
\end{figure}
\begin{figure}
	\centering
	\includegraphics[trim=40 0 10 0, width=8cm, height=5cm]{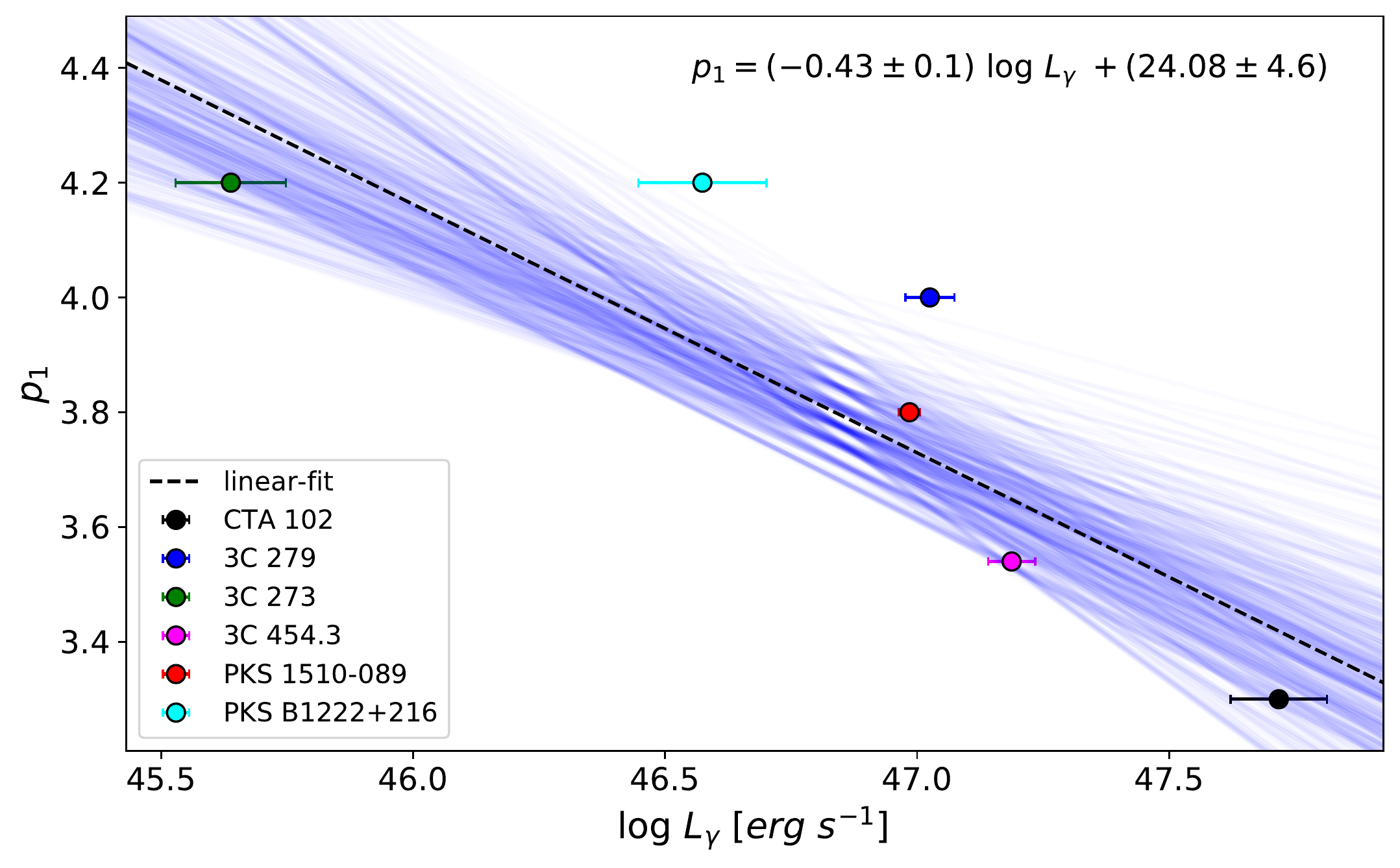}
	\caption{Dependence of intrinsic $\gamma$-luminosity (0.1 -- 300 GeV) on high energy slope of particle energy distibution ($p_1$) fitted with bootstrapped linear regression. Blue lines are bootstrapped regression lines and the black dotted line is the average of them.}
	\label{fig:gammalumvs-p1}
\end{figure}

\subsection{Comparison with earlier studies}
In this subsection the obtained results are compared with different results published earlier. For reference see \autoref{tab:Fermi}, \autoref{tab:SEDParameter} and \autoref{tab:EnergyPower}. \\
(i) \textbf{CTA 102:} \citet{2018ApJ...863..114G} studied the extreme outburst of 2016-17 and also the pre-flare low state. The low state and flaring states' \textit{Fermi}-LAT spectral indices are mentioned to be 2.39 and 2.01--1.81 respectively. They reported $P_e$$\sim$$10^{45}$ erg/s and $P_B$$\sim$$10^{46}$ erg/s in the low state, and $P_e$$\sim$$(10^{45}-10^{47})$ erg/s and $P_B$$\sim$$(10^{42}-10^{39})$ erg/s in the flaring states. In our quiescent state analysis, we obtained steeper \textit{Fermi}-LAT spectral index (2.5), lower $P_e$ ($\sim$$10^{44}$ erg/s) and similar $P_B$ ($\sim$$10^{46}$ erg/s) values as expected. \\
(ii) \textbf{3C 279:} Comparative study of low state, flares and post-flare state SED has been carried out by \citet{Pittori_2018}. The \textit{Fermi}-LAT spectral indices during two low states and two flares are (1.62, 1.77) and (1.53, 1.47) respectively. We obtained a steeper \textit{Fermi} index of 2.46. \citet{Pittori_2018} reported $P_e$$\sim$$10^{44}$ erg/s, $P_B$$\sim$$10^{44}$ erg/s and $P_p$$\sim10^{46}$ across all states, and $P_{rad}$$\sim$$10^{44}$, $10^{46}$ erg/s during low and high states respectively. We obtained higher $P_e$ and $P_B$ ($\sim$$10^{45}$ erg/s), similar $P_p$ but lower $P_{rad}$ ($\sim$$10^{44}$ erg/s) in quiescent state. \\
(iii) \textbf{3C 273:} \citet{2013A&A...557A..71R} did a comparative study of \textit{Fermi}-LAT spectrum during a low state and five different flare states. They reported spectral index of 2.86 during low state and 2.5--2.8 during the flare states. Broadband SED analysis of quiescent state is not yet done for this source. We got similar index of 2.9 during the selected quiescent state.\\
(iv) \textbf{3C 454.3:} \citet{2016ApJ...826...54D} did broadband SED modeling of an average state using data from \citet{2010ApJ...716...30A} and reported $P_e = 6.1\times10^{45}$ and $P_B = 1.18\times10^{45}$ erg/s, which are greater and less than values obtained by us respectively. \\
(v) \textbf{PKS 1510-089:} \citet{Acciari:2018ptm} did exclusive study of low-state broadband SED with MAGIC observations and reported spectral index of 2.56. Whereas we obtained spectral index of 2.41. They did not mention resulting powers carried by jet. So, we used their SED models to calculate jet powers for comparison purpose and obtained $P_e$, $P_B$ and $P_{rad}$ of the order of $10^{45}$, $10^{44}$ and $10^{44}$ erg/s respectively, which support our results. \\
(vi) \textbf{PKS B1222+216:}  \citet{10.1093/mnras/staa2958} carried out detailed comparison between flaring and quiescent states and reported \textit{Fermi} spectral indices of 2.5 during quiescence and 1.7--2.3 during flares, which is comparable to spectral index obtained in this work (2.6). They also did not mention resulting jet powers from SED modelings. \citet{2014ApJ79661K} reported total jet power of $2\times10^{46}$ erg/s during the brightest flare, which is less than what we obtained. \\
Thus, during quiescent states we get steeper \textit{Fermi}-LAT spectra as compared to flare states. Calculated jet power budgets during quiescent state in this work are comparable to the values mentioned in literature. But, there is no clear trend of variation of $P_e$ and $P_B$ with change in flux. Their changes depend on the type of the process used to describe the flux variation (eg. propagation of shock-front, geometric effects, magnetic reconnection etc).

\section{Discussion} 
\label{sec:dis}

(i) \textit{Black hole mass and disk luminosity:} It is worth noticing that in a quiescent state, thermal emission from accretion disk dominates over synchrotron radiation in the optical-UV band for all sources (except 3C 279). The trend shown by archival data underlying the actual analysed data in the optical-UV band indicates the presence of a big blue bump (BBB) which has a thermal origin. Thus, fitting the optical-UV band using thermal emission from multi-temperature blackbody type accretion disk helps to estimate disk luminosity and the central supermassive black hole mass. According to blazar sequence, generally synchrotron emission in FSRQs peak in the infrared band and contribute less in the optical-UV band. Thus, thermal emission from disk becomes visible in the form of BBB in SEDs. As the selected FSRQs are the brightest $\gamma$-ray emitters, they contain black holes with very large masses ($\ga 8\times10^8 M_{\sun}$) as expected. They also have very high mass accretion rate of (0.1--6) in Eddington units ($>$1$M_{\sun}$ yr$^{-1}$).\par
(ii) \textit{Emission region size:} The results listed in the previous section lead to the discussion on how the study of quiescent states of FSRQs is important. Though FSRQs are known for showing highly variable emission in all wavebands and occasional outbursts, they remain mostly in quiescent state during their lifetime. Previous studies on FSRQ flares have reported significant fast variability timescales of less than a day. Even timescales of the order of minutes have been reported in $\gamma$-ray waveband \citep{2018ApJ...854L..26S, Ackermann_2016, 2012ApJ...760...69N, Aleksi__2011}. But, from the \textit{Fermi}-lightcurves (\autoref{fig:alllc}) it can be seen that FSRQs show lower flux variability during quiescent states.  
The emission region sizes ($R$) of the order of 10$^{16}$ cm used for modeling quiescent SEDs (\autoref{tab:SEDParameter}) correspond to variability timescales of 1 to 5 days, which is larger than minute-hour variability timescales observed for these sources during flares. As expected, this implies slower variability in quiescence for all sources which corresponds to larger emission regions according to light travel-time argument.\par
(iii) \textit{Emission mechanisms:} According to conical jet model, larger emission regions were to be placed at larger distances from the central engines. Thus, emission regions for CTA 102, 3C 279, PKS 1510-089 and PKS B1222+216 were found to be outside BLR. Emission regions for 3C 273 and 3C 454.3 were found to be inside BLR. When $d < R_{BLR}$, seed photon contribution from BLR is maximum. High energy $\gamma$-ray data is then fitted by dominant $EC_{BLR}$ component with negligible contribution from $EC_{DT}$ component. When the emission region is placed outside BLR, $EC_{DT}$ component becomes important in describing $\gamma$-ray emission. In case of CTA 102 and PKS 1510-089 the emission regions were close to the outer edge of the BLR and far away from the dusty torus. That is why, despite being Doppler de-boosted in blob rest-frame, the EC components from BLR photon field ($EC_{BLR}$) dominate over EC component from torus photon field ($EC_{DT}$). In PKS B1222+216 and 3C 279, the emission regions are further away from BLR which indicate the dominance of $EC_{DT}$ over $EC_{BLR}$ as non-thermal high energy emission process. $\tau_{BLR}$ values for all sources were kept at 10\% except for CTA 102 ($\tau_{BLR}$ = 15\%), which is quite reasonable. A bit higher $\tau_{DT}$ value was required for 3C 279 to fit the SED. As dusty torus is a huge doughnut-shaped cloud surrounding the AGN and in 3C 279 it is quite close to the central engine due to less radiation pressure generated by low disk luminosity, we may assume that it covers larger solid angle to intercept a higher fraction of disk luminosity. Having the emission region placed outside BLR imply reduced photon field inside the emission region which makes the cooling process weaker. Thus, it is expected to have higher break energy in electron distribution where the difference between $d$ and $R_{BLR}$ is higher, i.e. the cooling process is weaker. \autoref{fig:gvsdrblr} shows that CTA 102, PKS 1510-089, PKS B1222+216 and 3C 279 follow this trend. Due to high variability in flare states, the emission region, being smaller in size, generally placed inside BLR. Thus, Doppler-boosted high $EC_{BLR}$ component is produced which can explain bright $\gamma$-ray flux.\par
(iv) \textit{Magnetic fields:} According to \citet{10.1111/j.1365-2966.2009.15898.x}, magnetic field in FSRQ jet is (1--10) G on an average. But we found that in quiescent state two sources (3C 279 and PKS 1510-089) have magnetic fields less than 1 G. From \autoref{fig:mbhvsb}, it seems that FSRQs containing more massive black holes have a higher magnetic field in the jet. But, this does not tell the whole picture. As the source of the magnetic field is the mass accretion process, we can expect a correlation between disk luminosity and magnetic field. The magnetic field inside the emission region should decrease with an increase in emission region distance from the central engine. Thus, all these effects together can explain the nice correlation (Pearson Correlation = -0.83) between \lq $B$' and \lq $\log_{10}(d/(R_S*L_d))$' shown in \autoref{fig:bvslogdld} as, $B$ = (-133 $\pm$ 38) - (3.2$\pm$0.9) $\log_{10}(d/(R_S*L_d))$. \par
(v) \textit{Particle energy distribution:} FSRQs have very powerful jets containing a highly dense population of charged particles. So, these particles get cooled very efficiently due to collision with each other. Because of this short cooling time-scale, injected particle distribution cannot get accelerated up to very high energy and cools down rapidly before escaping the emission region. So, $\gamma_{min}$ value for all sources were kept less than 10. Use of low $\gamma_{min}$ value yields very wide $EC_{BLR}$ or $EC_{DT}$ component which can simultaneously describe both X-ray and $\gamma$-ray data in flare state \citep{2011A&A...530A...4A}. But, in quiescent state SEDs, a significant contribution of SSC or $EC_{disk}$ component is required depending on the distance of emission region from the central engine to fit low energy X-ray data along with $EC_{BLR}$ or $EC_{DT}$. Electron densities ($N_e$) found are quite high ($\sim$$10^3$--$10^4$ cm$^{-3}$) which imply powerful jets. We used electron energy distribution of broken power-law shape. This kind of particle distribution can be produced by underlying Fermi first and second-order particle acceleration processes \citep{2019hepr.confE..75L}. The slopes of both synchrotron and EC component above peak frequencies are dependent on the high energy index of particle energy distribution ($p_1$). For all sources, \lq $p_1$' value was found to be more than 3, which implies that electrons having energy around the break in energy distribution radiate the maximum amount of energy. Higher $p_1$ value implies less number of radiating electrons at higher energies which in turn result in lower $\gamma$-ray luminosity. As expected, the $\gamma$-ray luminosity observed by \textit{Fermi}-LAT was found to be decreasing (Pearson Correlation = -0.85) with $p_1$ as, $p_1$ = (24$\pm$4) - (0.4 $\pm$ 0.1)$\log_{10}L_{\gamma}$ (\autoref{fig:gammalumvs-p1}). We found $\gamma_{max}$ values of the order of $\sim$($10^3$--$10^4$). To describe flare SEDs, higher $\gamma_{max}$ and $\gamma_{break}$ values and harder particle spectrum are required \citep{2016ApJ...832...17B, 2014ApJ...790...45P, 2017A&A...603A..29A}.\par
(vi) \textit{Power budget:} Radiated power of only two sources (CTA 102 and 3C 454.3) are higher than the powers carried by electrons. Considering one proton per ten electrons, kinetic power carried by protons ($P_p$) is found to be much higher than $P_e$ in all sources. Even in quiescent states, Poynting flux carries a significant amount of power ($P_B \ga P_e$). Thus, even though power carried by electrons is not always sufficient to account for the observed radiation power, proton kinetic power and Poynting flux power are always individually sufficient in a quiescent state. \citet{10.1111/j.1365-2966.2009.15898.x} reported that, for FSRQs in average state $P_{rad}$ is greater than $P_e$. But, in the quiescent state, we can see that most of the brightest FSRQs have sufficient kinetic power carried by electrons to account for observed radiation. Jets can be powered by mass accretion process and the black hole spin. To explain the $P_{jet} \ge L_d$ scenario, \citet{10.1111/j.1365-2966.2009.15554.x} proposed that a large fraction of accretion power is spent to power the jet and a smaller fraction ($\epsilon \sim 0.08$) produces disk luminosity. \citet{10.1111/j.1365-2966.2009.15898.x} mentioned that we can have $P_{jet} \ge L_d$ if there is an efficient way to extract energy from the spinning central black hole. According to the Blandford-Znajek (BZ) mechanism of extracting the energy of a spinning black hole \citep{1977MNRAS.179..433B}, $P_{jet}$ is proportional to $L_d$ where the value of proportionality constant can be greater than unity. \par
(vii) \textit{Application:} FSRQs mostly remain in a state of low activity during their lifetime. So, the quiescent state SED models represent the underlying consistent emission process in the FSRQs. The high activity states can be considered as perturbations of some physical parameters in the jet environment or emergence of any additional emission region. Thus, for modeling brighter and more variable states of the FSRQs selected in this work, emission from a smaller region should be added to the quiescent state SED model while fitting the SED data points. Morever, time-dependant leptonic models can be used to describe the evolution of flares from the quiescent state \citep{2014JHEAp...1...63D}. The quiescent SED model parameters can be used as the starting point of the time-evolution process. The observed cross-band correlations and associated time-lags can be explained as perturbations of some of those initial parameters.
\section{Conclusion}
\label{sec:con}

After extensive study of quiescent state spectral energy distributions of six brightest FSRQ detected by
\textit{Fermi}-LAT, we have arrived at the following conclusions. \\
(i) For most of the FSRQs in quiescence, low synchrotron emission is dominated by thermal emission from accretion disk in optical-UV waveband. Thus, disk luminosity and black hole mass can be estimated using \textit{Swift}-UVOT data. Estimated black hole masses and disk luminosities are consistent with values mentioned in literature (\autoref{tab:sourcename} and \autoref{tab:SEDParameter}).\\
(ii) Slow variability of $>$ 1-day timescale in the quiescent state indicates a large emission region size.\\
(iii) Following conical jet structure and assuming that the emission regions fill the whole jet cross-section, the large emission regions were found to be mostly outside BLR. Thus, low $\gamma$-ray flux in quiescent state can be described by Doppler-de-boosted (in blob rest frame) weak $EC_{BLR}$ component and Doppler-boosted (in blob rest frame) weak $EC_{DT}$ component (as $R_{DT} \simeq 25 R_{BLR}$) depending on emission region distance from BLR.\\
(iv) In most of the quiescent FSRQ SED models, a small bump is clearly visible in the IR waveband due to the contribution of thermal emission from dusty torus. If nearly simultaneous mm-IR observations were available, it would have been possible to constrain torus temperature and the fraction of disk luminosity intercepted by the dusty torus. \\
(v) In the quiescent state, blazar jets become radiatively inefficient. In most of the brightest FSRQs, the kinetic power carried by electrons can account for observed radiation. But still, the jets remain powerful enough to radiate more than disk luminosity.\\
(vi) We could not constrain the synchrotron peak positions properly in the SED models due to lack of simultaneous mm-IR observations. So we can roughly say, in quiescent states of CTA 102, 3C 279 and 3C 454.3 synchrotron and IC emission peak at lower frequencies than those in flare states.\\
(vii) Magnetic field inside the emission region has an expected correlated with \lq $d/(R_S*L_d)$' for these six sources during quiescent state. \\
(viii) The high energy indices of particle spectrum ($p_1$) of these six sources are found to be correlated with quiescent state $\gamma$-ray luminosities ($L_{\gamma}$) observed by \textit{Fermi}-LAT in 0.1--300 GeV energy range. For FSRQs, \textit{Fermi}-LAT data form the decaying part of the high energy hump in SED, i.e. $L_{\gamma}$ represents the luminosity emitted by particles with higher energies. Higher $p_1$ value implies lower number of high energy particles (\autoref{eq:3}), which in turn reduces $L_{\gamma}$.

\section*{Acknowledgements}

We would like to thank the anonymous referee whose valuable suggestions truly helped to improve our work.
This research has made use of data from the OVRO 40-m monitoring program (Richards, J. L. et al. 2011, ApJS, 194, 29) which is supported in part by NASA grants NNX08AW31G, NNX11A043G, and NNX14AQ89G and NSF grant AST-0808050 and AST-1109911. We are grateful to Sebastian Kiehlmann for providing us non-public OVRO lightcurves of some sources.
Data from the Steward Observatory spectropolarimetric monitoring project were used. This program is supported by Fermi Guest Investigator grants NNX08AW56G, NNX09AU10G, NNX12AO93G, and NNX15AU81G.
This research has made use of the XRT Data Analysis Software (XRTDAS) developed under the responsibility of the Space Science Data Center (SSDC) managed by Italian Space Agency (ASI).
Data obtained at the Mets\"{a}hovi Radio Observatory operated by the Aalto University were also used in this publication.
We have used Fermi-LAT data, obtained from the Fermi Science Support Center, provided by NASA's Goddard Space Flight Center (GSFC). The data and analysis software were obtained from NASA's High Energy Astrophysics Science Archive Research Center (HEASARC), a service of GSFC. We used a community-developed Python package named \textit{Enrico} to make \textit{Fermi}-LAT data analysis easier and more convenient \citep{2013ICRC...33.2784S}.
Finally, We acknowledge the support of the Department of Atomic Energy, Government
of India, under Project Identification No. RTI 4002.

\section{DATA AVAILABILITY STATEMENT}
OVRO data is available at: \url{https://www.astro.caltech.edu/ovroblazars/}.\\
Steward Observatory SPOL-CCD data is available at: \url{http://james.as.arizona.edu/~psmith/Fermi/}. \\
SMARTS data is available at: \url{http://www.astro.yale.edu/smarts/glast/home.php}. \\
\textit{Swift}-mission data is available at: \url{https://heasarc.gsfc.nasa.gov/docs/archive.html}. \\
\textit{Fermi}-LAT data is available at: \url{https://fermi.gsfc.nasa.gov/ssc/data/access/}. \\
SSDC (ASI) archive: \url{https://www.ssdc.asi.it/}. \\
\textit{Swift}-mission data analysis software package is available at: \url{https://heasarc.gsfc.nasa.gov/lheasoft/download.html}. \\
\textit{Fermi}-LAT data analysis software is available at: \url{https://fermi.gsfc.nasa.gov/ssc/data/analysis/software/}. \\
Software for Broadband SED modeling is available at: \url{https://github.com/andreatramacere/jetset}.





\bibliography{QSBB}
\bibliographystyle{mnras}

\bsp	
\label{lastpage}
\end{document}